\documentclass[twocolumn,showpacs,aps,prl,superscriptaddress]{revtex4}

\usepackage{graphicx}
\usepackage{dcolumn}
\usepackage{amsmath}
\usepackage{epsfig}

\RequirePackage{xspace}

\usepackage{relsize}
\def\babar{\mbox{\slshape B\kern-0.1em{\smaller A}\kern-0.1em
    B\kern-0.1em{\smaller A\kern-0.2em R}}}

\def\epem       {\ensuremath{e^+e^-}\xspace}

\def\pip   {\ensuremath{\pi^+}\xspace}
\def\pim   {\ensuremath{\pi^-}\xspace}

\def\Kbar  {\kern 0.2em\overline{\kern -0.2em K}{}\xspace}

\def\Kz    {\ensuremath{K^0}\xspace}
\def\Kzb   {\ensuremath{\Kbar^0}\xspace}
\def\KzKzb {\ensuremath{\Kz \kern -0.16em \Kzb}\xspace}
\def\Kp    {\ensuremath{K^+}\xspace}
\def\Km    {\ensuremath{K^-}\xspace}

\def\KpKm  {\ensuremath{\Kp \kern -0.16em \Km}\xspace}

\def\Dbar    {\kern 0.2em\overline{\kern -0.2em D}{}\xspace}

\def\Dz      {\ensuremath{D^0}\xspace}
\def\Dzb     {\ensuremath{\Dbar^0}\xspace}
\def\DzDzb   {\ensuremath{\Dz {\kern -0.16em \Dzb}}\xspace}
\def\Dp      {\ensuremath{D^+}\xspace}
\def\Dm      {\ensuremath{D^-}\xspace}

\def\DpDm    {\ensuremath{\Dp {\kern -0.16em \Dm}}\xspace}

\def\Bbar    {\kern 0.18em\overline{\kern -0.18em B}{}\xspace}

\def\BB      {\ensuremath{B\Bbar}\xspace} 
\def\Bz      {\ensuremath{B^0}\xspace}
\def\Bzb     {\ensuremath{\Bbar^0}\xspace}
\def\BzBzb   {\ensuremath{\Bz {\kern -0.16em \Bzb}}\xspace}
\def\Bu      {\ensuremath{B^+}\xspace}
\def\Bub     {\ensuremath{B^-}\xspace}

\def\BpBm    {\ensuremath{\Bu {\kern -0.16em \Bub}}\xspace}

\def\BorBbar    {\kern 0.18em\optbar{\kern -0.18em B}{}\xspace}
\def\DorDbar    {\kern 0.18em\optbar{\kern -0.18em D}{}\xspace}
\def\KorKbar    {\kern 0.18em\optbar{\kern -0.18em K}{}\xspace}

\mathchardef\Upsilon="7107
\def\Y#1S{\ensuremath{\Upsilon{(#1S)}}\xspace}

\mathchardef\Deltares="7101
\mathchardef\Xi="7104
\mathchardef\Lambda="7103
\mathchardef\Sigma="7106
\mathchardef\Omega="710A

\def\Deltabar{\kern 0.25em\overline{\kern -0.25em \Deltares}{}\xspace}
\def\Lbar{\kern 0.2em\overline{\kern -0.2em\Lambda\kern 0.05em}\kern-0.05em{}\xspace}
\def\Sigbar{\kern 0.2em\overline{\kern -0.2em \Sigma}{}\xspace}
\def\Xibar{\kern 0.2em\overline{\kern -0.2em \Xi}{}\xspace}
\def\Obar{\kern 0.2em\overline{\kern -0.2em \Omega}{}\xspace}
\def\Nbar{\kern 0.2em\overline{\kern -0.2em N}{}\xspace}
\def\Xb{\kern 0.2em\overline{\kern -0.2em X}{}\xspace}

\def\BR         {{\ensuremath{\cal B}\xspace}}

\def\mes        {\mbox{$m_{\rm ES}$}\xspace}

\newcommand{\tev}{\ensuremath{\mathrm{\,Te\kern -0.1em V}}\xspace}
\newcommand{\gev}{\ensuremath{\mathrm{\,Ge\kern -0.1em V}}\xspace}
\newcommand{\mev}{\ensuremath{\mathrm{\,Me\kern -0.1em V}}\xspace}
\newcommand{\kev}{\ensuremath{\mathrm{\,ke\kern -0.1em V}}\xspace}
\newcommand{\ev}{\ensuremath{\mathrm{\,e\kern -0.1em V}}\xspace}
\newcommand{\gevc}{\ensuremath{{\mathrm{\,Ge\kern -0.1em V\!/}c}}\xspace}
\newcommand{\mevc}{\ensuremath{{\mathrm{\,Me\kern -0.1em V\!/}c}}\xspace}
\newcommand{\gevcc}{\ensuremath{{\mathrm{\,Ge\kern -0.1em V\!/}c^2}}\xspace}
\newcommand{\mevcc}{\ensuremath{{\mathrm{\,Me\kern -0.1em V\!/}c^2}}\xspace}

\def\invfb   {\ensuremath{\mbox{\,fb}^{-1}}\xspace}

\def\mus  {\ensuremath{\rm \,\mus}\xspace}

\def\mus        {\ensuremath{\,\mu{\rm s}}\xspace}    

\def\to                 {\ensuremath{\rightarrow}\xspace}

\def\pep2{PEP-II}

\def\gsim{{~\raise.15em\hbox{$>$}\kern-.85em
          \lower.35em\hbox{$\sim$}~}\xspace}
\def\lsim{{~\raise.15em\hbox{$<$}\kern-.85em
          \lower.35em\hbox{$\sim$}~}\xspace}

\xspace

\newcommand{\jprlBase}       {Phys.\ Rev.\ Lett.\xspace}
\newcommand{\jprBase}        {Phys.\ Rev.\xspace}
\newcommand{\jplBase}        {Phys.\ Lett.\xspace}
\newcommand{\nimBaseA}       {Nucl.\ Instr.\ Meth.\xspace}

\newcommand{\npBase}         {Nucl.\ Phys.\xspace}

\newcommand{\nima}      [1]  {\nimBaseA~A~{\bf #1}}

\newcommand{\npb}       [1]  {\npBase\ B~{\bf #1}}

\newcommand{\plb}       [1]  {\jplBase\ B~{\bf #1}}

\newcommand{\jprl}      [1]  {\jprlBase\ {\bf #1}}
\newcommand{\jprd}      [1]  {\jprBase\ D~{\bf #1}}

\def\jetset74   {\mbox{\tt Jetset \hspace{-0.5em}7.\hspace{-0.2em}4}\xspace}

\def\de {\ensuremath{\Delta E}}

\def\figurebox#1#2#3{%
    \def\arg{#3}%
    \ifx\arg\empty
    {\hfill\vbox{\hsize#2\hrule\hbox to #2{\vrule\hfill\vbox to 
#1{\hsize#2\vfill}\vrule}\hrule}\hfill}%
    \else
    {\hfill\epsfbox{#3}\hfill}%
    \fi}

\long\def\inst#1{\par\nobreak\kern 4pt\nobreak
    {\it #1}\par\vskip 10pt plus 3pt minus 3pt}

\begin{document}

\begin{flushleft}
\babar-PUB-04/003 \\
SLAC-PUB-10367\\
\end{flushleft}

\title{
{
\Large \bf \boldmath 
Search for the Decay $\Bz\to p\bar{p}$
}
}

%
\author{B.~Aubert}
\author{R.~Barate}
\author{D.~Boutigny}
\author{F.~Couderc}
\author{J.-M.~Gaillard}
\author{A.~Hicheur}
\author{Y.~Karyotakis}
\author{J.~P.~Lees}
\author{V.~Tisserand}
\author{A.~Zghiche}
\affiliation{Laboratoire de Physique des Particules, F-74941 Annecy-le-Vieux, France }
% ...
\collaboration{The \babar\ Collaboration}
\noaffiliation



\begin{abstract}
We present the result of a search for the charmless two-body baryonic decay 
$B^0\rightarrow p\bar{p}$ in a sample of $88$ million 
$\Upsilon(4{\rm S})\rightarrow B\bar{B}$ decays collected 
by the BaBar detector at the SLAC PEP-II asymmetric-energy $B$ Factory. We
use Cherenkov radiation to identify protons cleanly, and determine the signal 
yield with a maximum-likelihood fit technique using kinematic and topological 
information. We find no evidence for a signal and place a $90\%$ 
confidence-level upper limit of 
$\BR(B^0\rightarrow p\bar{p}) < 2.7\times 10^{-7}$.  

\end{abstract}


\maketitle

\setcounter{footnote}{0}


We report the result of a search for the charmless two-body baryonic decay 
$\Bz\to p\bar{p}$~\cite{ref:cc}.  Although $B$ mesons have recently been
observed to decay into several charmless three-body baryonic final 
states~\cite{ref:ppbarK}, there is currently no evidence for the 
corresponding charmless two-body decays.  
Previous searches~\cite{belleppbar,cleoppbar} for $\Bz\to p\bar{p}$ decays
have yielded upper limits on the branching fraction at the level of $10^{-6}$, which is 
consistent with calculations based on QCD sum rules~\cite{ref:QCDsumrule} and the 
pole model~\cite{ref:PoleModel}.  
A simple scaling of the measured branching fraction for 
$\Bz\to \Lambda_c^- p$~\cite{ref:Lambdacp} 
by the current estimate~\cite{ref:PDG2002} of 
$\left |V_{\rm ub}/V_{\rm cb}\right |^2$ leads to a prediction of charmless 
two-body branching fractions at the level of $10^{-7}$, which is near the 
current sensitivity of present experiments.

The data sample used for this search contains $(87.9\pm 1.0)\times 10^6$ 
$\Y4S\to\BB$ decays collected by the \babar\ detector~\cite{ref:babar} at the 
SLAC PEP-II $\epem$ asymmetric-energy storage ring.  The primary detector 
components used in the analysis are a charged-particle tracking system 
consisting of a five-layer silicon vertex detector and a 40-layer drift chamber 
surrounded by a $1.5$-T solenoidal magnet, and a dedicated particle 
identification system consisting of a detector of internally reflected 
Cherenkov light (DIRC).

Two-body $B$ decays are reconstructed from pairs of 
oppositely-charged tracks originating from the interaction region and 
having momentum greater than $100\mevc$ in the direction transverse to the
beam line.  We require each track to have an associated Cherenkov angle 
($\theta_c$) measurement with at least four signal 
photons detected in the DIRC.  To suppress combinatorial background arising 
from $\Lambda$ decays, we require that the two tracks form a vertex with 
probability greater than $10^{-3}$.

Signal candidates are identified kinematically with two variables: the difference 
$\de$ between the center-of-mass (CM) energy of the $B$ candidate and 
$\sqrt{s}/2$, where $\sqrt{s}$ is the total CM energy, and the beam-energy 
substituted mass 
$\mes = \sqrt{(s/2 + {\mathbf {p}}_i\cdot {\mathbf {p}}_B)^2/E_i^2- {\mathbf 
{p}}_B^2}$, with the $B$-candidate momentum 
${\mathbf {p_B}}$ and the four-momentum of the initial state 
$(E_i, {\mathbf {p_i}})$ defined in the laboratory frame.
For signal decays, $\de$ peaks near zero with a resolution of 
about $23\mev$, while $\mes$ peaks near the $B$ mass with a 
resolution of about $2.6\mevcc$.  We require $5.20 < \mes < 5.29\gevcc$ and 
$\left|\de\right|<100\mev$.

\begin{figure}[!htb]
\begin{center}
\includegraphics[height=7cm]{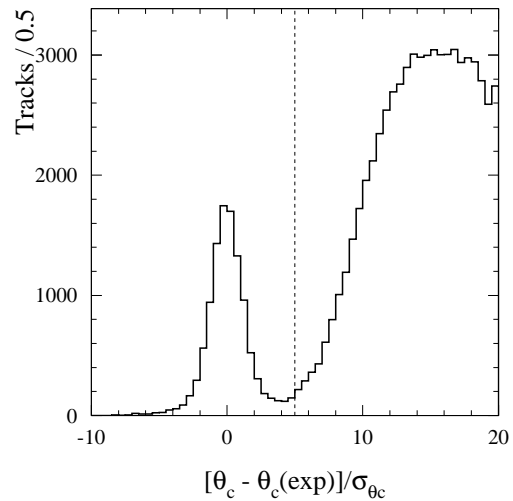}
\caption{The difference between the measured and expected values
of $\theta_c$, divided by the error, for tracks from $\Bz\to p\bar{p}$ 
candidates in the region $5.20 < \mes < 5.26\gevcc$.  We only use
tracks that lie on the left side of the dashed line.}
\label{fig:CherenkovPull}
\end{center}
\end{figure}

Protons are identified based on the $\theta_c$ measurement from the DIRC and
the momentum measurement from the tracking system.  
Figure~\ref{fig:CherenkovPull} 
shows the difference between measured and expected values of $\theta_c$ for
the proton hypothesis, divided by the error $\sigma_{\theta_c}$, for tracks 
from $\Bz\to p\bar{p}$ candidates in the sideband region 
$5.20 < \mes < 5.26\gevcc$.  
Protons peak near zero and are well separated from the much larger background 
of pions and kaons.  We require a $\theta_c$ measurement within 
$5\sigma_{\theta_c}$ of the expected value for a proton, which removes over 
$97\%$ of the combinatorial background while retaining more than $91\%$ of the 
signal decays (the efficiency is less than $100\%$ due to the
presence of non-gaussian tails in the pull distribution).  

We measure the efficiency of the $\theta_c$ selection in a sample of
$\Lambda \to p\pi^-$ decays reconstructed in $9.6\invfb$ of $\epem$
annihilation data recorded $40\mev$ below the $\Y4S$ resonance.
The sample is selected using kinematic and decay-vertex information, and
has a purity of $98.5\%$.  For consistency with $\Bz\to p\bar{p}$ 
decays, we require the proton CM momentum $p^*$ to be in the range
$2.2 < p^* < 2.8\gevc$.  

Due to the unique topology and kinematics of the two-body final state, 
$b\to c$ decays do not populate the signal region for $\Bz\to p\bar{p}$,
and backgrounds from $b\to u$ decays are negligible after the proton 
selection.  We verify both assertions by analyzing a sample of approximately 
$80\invfb$ of $\Y4S\to\BB$ Monte Carlo simulated events in which the 
$B$ mesons decay according to the world-average branching 
fractions~\cite{ref:PDG2002}, and a second sample corresponding to 
approximately $200\invfb$ where one $B$ meson in each event is forced to 
decay to a charmless final state.  No event passes the above selection 
requirements in either sample.

The dominant background is from random combinations of protons produced in
the process $\epem\to q\bar{q}$ ($q=u,d,s$).  We verify in Monte Carlo
samples that the background from the process $\epem\to c\bar{c}$ is 
negligible compared to light-quark production.  In contrast to the
spherical topology of $\BB$ events, particles produced in light-quark events
tend to lie near the thrust axis of the original $q\bar{q}$ pair.  To 
suppress this background, we calculate the angle $\theta_S$ between the
sphericity axis of the $B$ candidate and the sphericity axis of the remaining
particles in the event, and require $\left |\cos{\theta_S}\right | < 0.9$.
This requirement removes $70\%$ of the combinatorial background, while retaining
$85\%$ of the signal decays.  In addition, we define a Fisher discriminant 
${\cal F}$~\cite{ref:pipiPRL}, which is a sum of two discriminating
variables with coefficients optimized to separate signal and light-quark 
events.  The first variable is the scalar sum of the CM momenta of all
the particles in the event, excluding the two tracks from the 
$\Bz\to p\bar{p}$ candidate.  The second variable is the product 
$p^*\left( \cos{\theta^*}\right)^2$ summed over all particles (excluding
the $B$-candidate tracks), where $\theta^*$ is the angle between its 
momentum and the $B$-candidate thrust axis in the CM frame.

The total efficiency for all of the above selection criteria is 
$(35.8\pm 3.7)\%$, where the error includes the statistical and
systematic uncertainties added in quadrature.  The dominant source
of the uncertainty is due to the limited statistical precision of the
$\Lambda$ control sample after applying the proton $p^*$ constraint.
A total of $804$ events satisfy the $\Bz\to p\bar{p}$ selection criteria.

The signal yield is determined from a maximum likelihood fit that uses
$\mes$, $\de$, and ${\cal F}$ as discriminating variables.  
The likelihood for the sample is defined as
\begin{equation}
{\cal L} = e^{-(N_{\rm S} + N_{\rm B})}\prod_{i=1}^{N}\left[N_{\rm S}{\cal P}^i_{\rm S} 
+ N_{\rm B}{\cal P}^i_{\rm B} \right],
\end{equation}
where $N$ is the total number of events in the sample, 
$N_{\rm S}$ and $N_{\rm B}$ are the signal (S) and background (B) yields, and 
${\cal P}^i_{\rm S}$ and ${\cal P}^i_{\rm B}$ are the signal and background 
probability density functions (PDFs) evaluated for event $i$.  The PDFs are 
calculated from the product of PDFs for the individual variables, which are 
taken to be uncorrelated in the fit.  We verify this assumption by calculating 
the linear correlation coefficient for each pair of variables.  The largest 
correlation ($-13\%$) is between $\mes$ and $\de$ in signal decays, and we
have confirmed that the effect of this small correlation is negligible.  The 
signal yield is determined by minimizing the function $-2\ln{\cal L}$ with 
respect to $N_{\rm S}$ and $N_{\rm B}$.

We use data and Monte Carlo samples to model the PDF shapes for signal
decays.  The mean and resolution of $\mes$ are dominated by the beam energy, 
and are therefore similar in decay modes where the momentum 
resolution of the $B$ candidate is significantly better than the resolution on 
the beam energy.  We obtain the mean and resolution of $\mes$, and also
the mean of $\de$, from a large sample of $\Bub\to\Dz(\Km\pip)\pim$ decays 
reconstructed in data.  Due to the difference in momentum resolution between 
protons and mesons, the resolution on $\de$ is different for $p\bar{p}$ and 
$\Dz\pim$ decays.  We therefore use the value obtained in a large Monte Carlo 
sample of $\Bz\to p\bar{p}$ decays, and apply a $5\%$ correction to account for the 
observed difference in $\de$ resolution for $\Dz\pim$ decays reconstructed
in data and Monte Carlo samples.
For ${\cal F}$ we use an asymmetric gaussian function with parameters obtained 
from simulated events.  The shapes of the background PDFs are obtained from 
data in the sideband regions $100 <\left|\de \right| < 200\mev$ and 
$5.20 < \mes < 5.26\gevcc$.  We use a linear shape for $\de$, a double-gaussian 
function for ${\cal F}$, and an empirical threshold function for 
$\mes$~\cite{ref:argus}.

Several cross-checks are performed to validate the fitting technique.  To 
confirm that the signal yield is unbiased, we generate and fit a large set of 
pseudo-experiments where signal and background events are generated randomly
from the PDFs.  For these studies, we assume a branching fraction of 
$10^{-6}$ and find that the fitted signal yield is unbiased.  We also 
check for biases arising from kinematic correlations by mixing Monte Carlo 
signal events with backgrounds generated directly from the PDFs.  No 
significant biases are observed.  The sensitivity of the analysis is determined 
from a set of pseudo-experiments with assumed branching fractions in
the range $(0.1$-$1.2)\times 10^{-6}$.  We find that for any branching fraction 
above $0.5\times 10^{-6}$, the null hypothesis would be excluded with 
a probability greater than $99.997\%$, corresponding to a significance
of $5\sigma$ for a gaussian distribution.

The result of the fit is $N_{\rm S} = -0.3^{+3.1}_{-2.0}$, consistent with no 
signal.  We determine a Bayesian $90\%$ confidence-level (C.L.) upper limit 
on $N_{\rm S}$ by finding the value $N_{\rm S}^{\rm UL}$ such that
\begin{equation}
\frac{\int_0^{N_{\rm S}^{\rm UL}}{\cal L}_{\rm max}dN_{\rm S}}
{\int_0^{\infty}{\cal L}_{\rm max}dN_{\rm S}}=0.90,
\end{equation}
where ${\cal L}_{\rm max}$ is the value of the likelihood as a function of 
$N_{\rm S}$.  We find $N_{\rm S}^{\rm UL}= 6.3$ events.

\begin{figure}[!htb]
\begin{center}
\includegraphics[height=4.25cm]{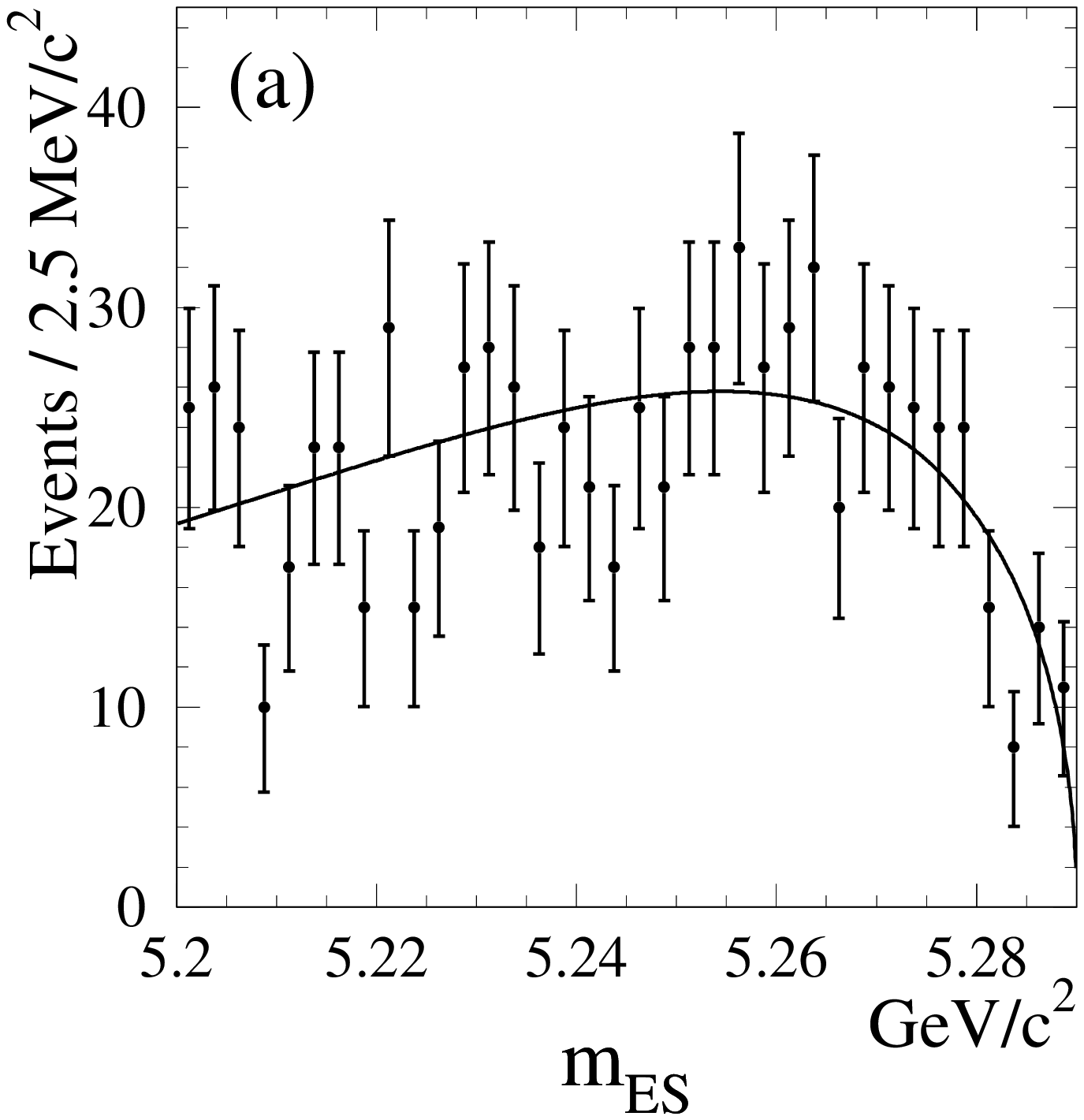}
\includegraphics[height=4.25cm]{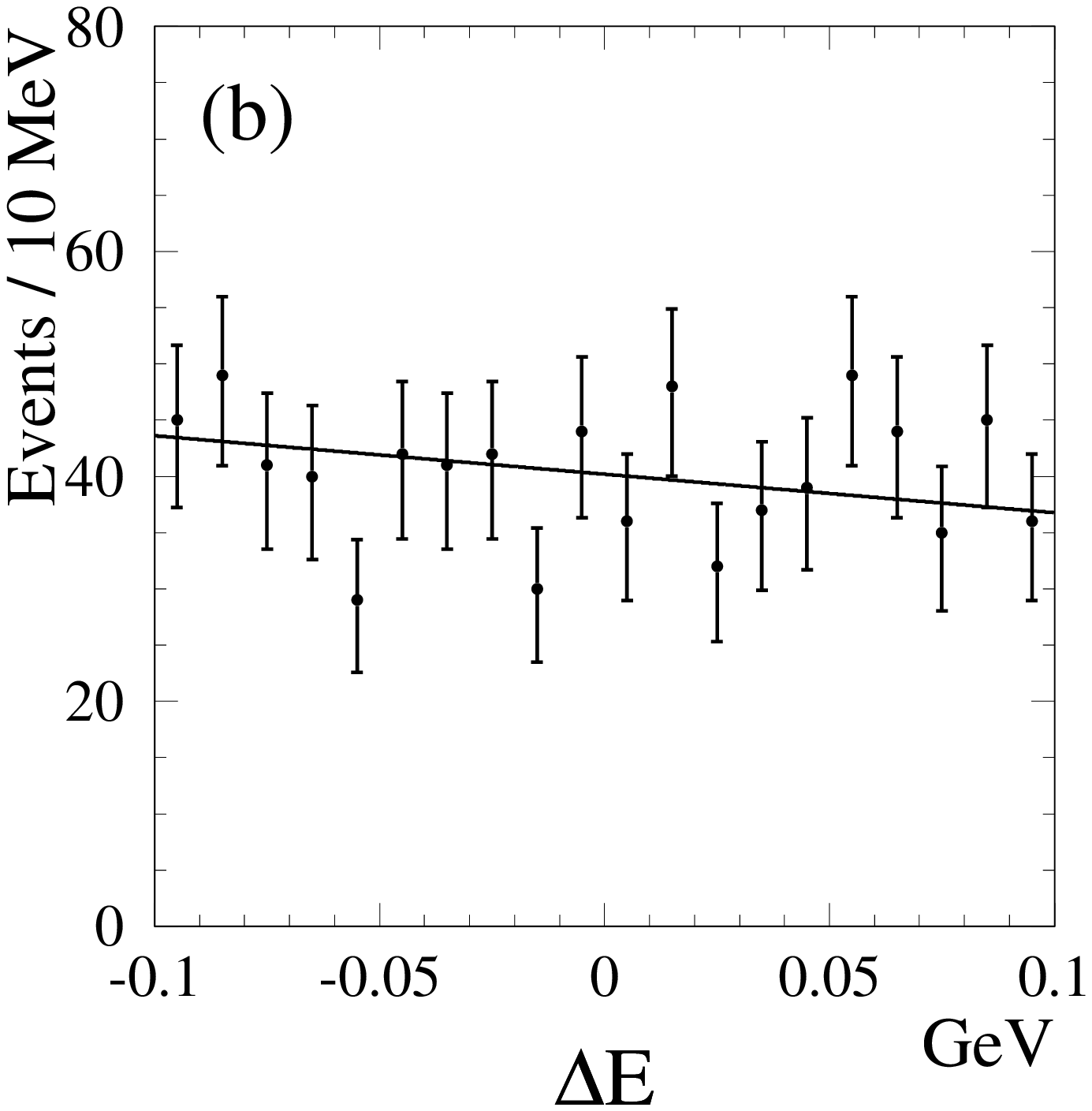}
\includegraphics[height=4.25cm]{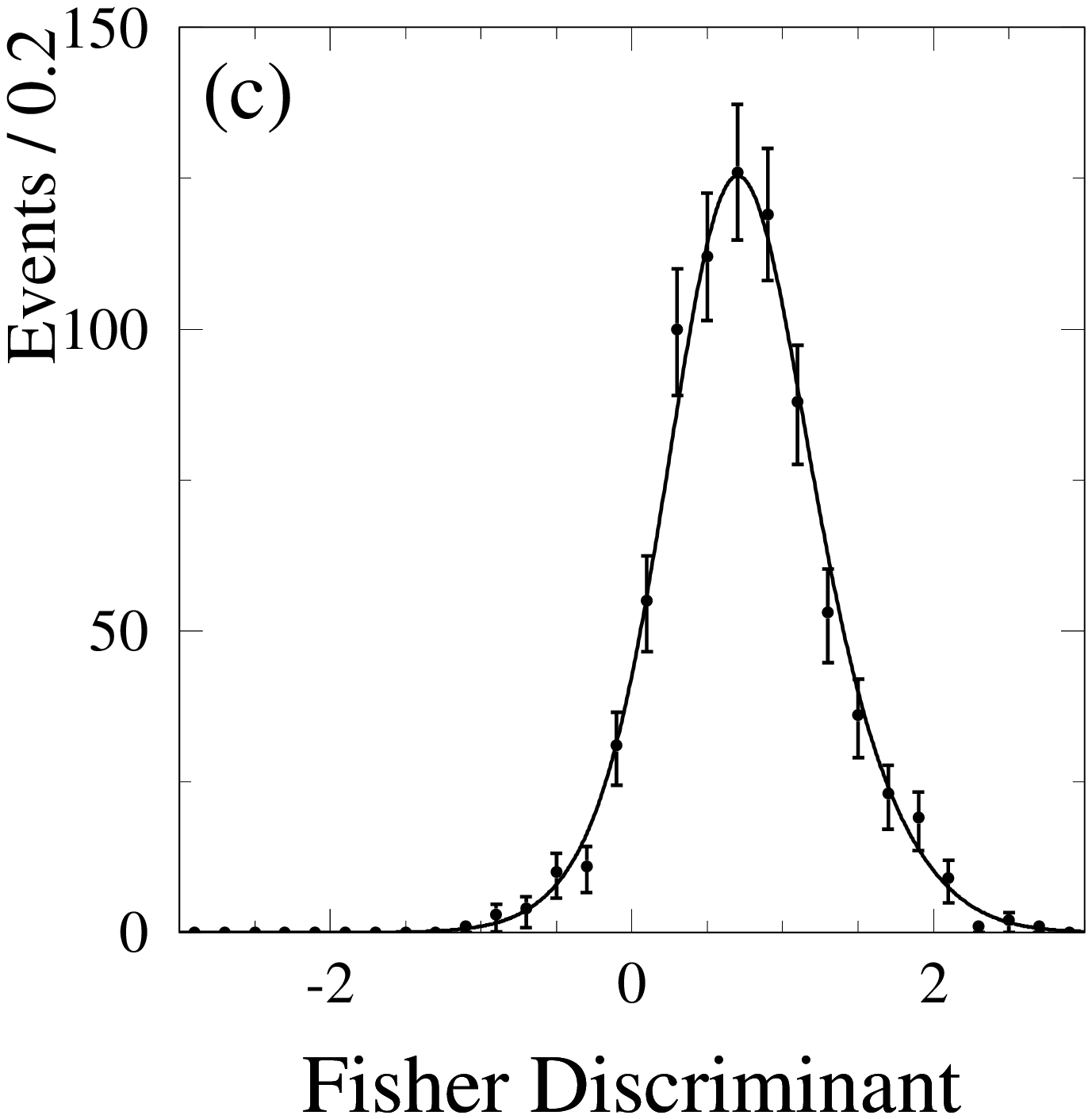}
\includegraphics[height=4.25cm]{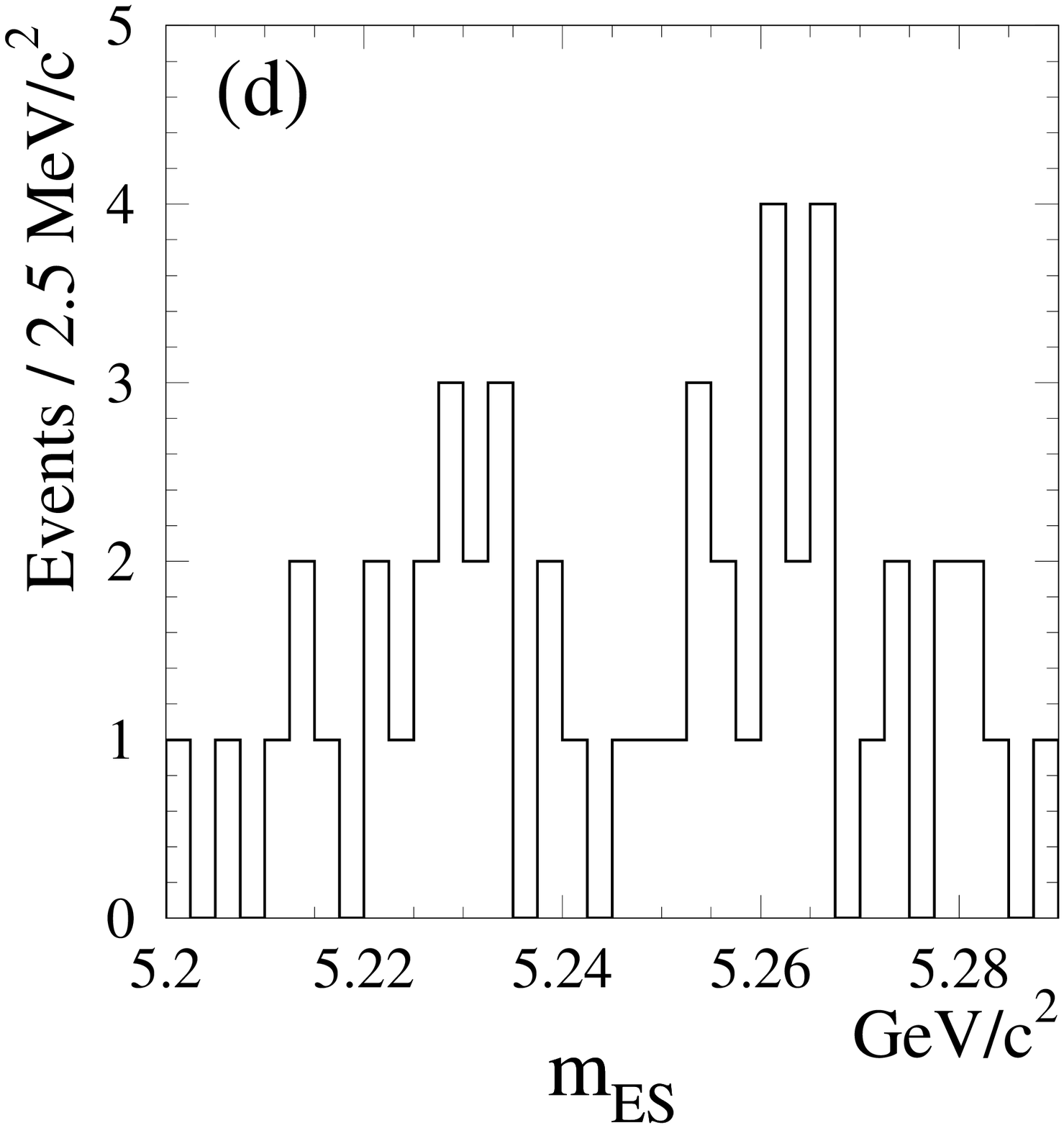}
\caption{Data (points with errors) and the result of the maximum likelihood fit 
(solid line) projected onto the (a) $\mes$, (b) $\de$, and (c) ${\cal F}$ variables, 
and (d) the distribution of $\mes$ for events in the signal-enhanced sample defined 
by the requirements $\left |\cos{\theta_S}\right | < 0.7$ and 
$\left |\de \right | < 30\mev$.}
\label{fig:results}
\end{center}
\end{figure}

Figures~\ref{fig:results}(a-c) show projections of the fit result in each of 
the discriminating variables.  The data in the signal region agree well with 
the background PDF shapes determined from sideband data.  As a cross-check on 
the fit, we apply more stringent background-rejection criteria and determine
the signal yield from the observed number of events in a restricted signal
region in $\mes$ and $\de$.  We require 
$\left |\cos{\theta_S}\right | < 0.7$, and define the signal region as 
$\left |\de\right | < 30\mev$ and $5.27< \mes < 5.29\gevcc$.  
Figure~\ref{fig:results}(d) shows the $\mes$ distribution for events passing the 
more restrictive $\cos{\theta_S}$ and $\de$ requirements.  There are $9$ events 
in the signal region with an expected background of $7.7\pm 1.4$, where the 
background is determined by extrapolating the observed yield in the sideband 
region $5.2 <\mes < 5.26\gevcc$.  The signal yield is $1.3\pm 3.3$, which is 
consistent with the null result from the likelihood fit.

Tables~\ref{tab:signalyield} and~\ref{tab:efficiency} summarize the 
various sources of systematic error on the signal yield and efficiency.  
Systematic uncertainty on $N_{\rm S}$ may arise from imperfect knowledge of 
the PDF parameters.  We vary each parameter by its estimated error and combine 
in quadrature the resulting variations in $N_{\rm S}$.  For the efficiency of 
the proton selection, we assign the 
$1.5\%$ background fraction in the $\Lambda$ control sample as the correlated 
systematic error.  The efficiency of the vertex quality requirement is 
determined to be $97.5\%$ from simulated $p\bar{p}$ decays, and we assign a 
systematic uncertainty of $2.5\%$ to account for possible differences between 
data and Monte Carlo events.  As a cross-check, we compare the efficiency in 
the topologically similar decays $\Bz\to\pip\pim$ and $\Bz\to\Kp\pim$ and find 
good agreement between data and simulation.  Finally, we include a correlated 
systematic error of $0.8\%$ per track to account for possible differences in 
tracking efficiency between data and Monte Carlo events.  The total systematic 
uncertainty on the efficiency is computed by adding correlated errors linearly, 
and then adding the separate sources in quadrature.

\begin{table}[!htb]
\caption{Summary of absolute systematic uncertainties on $N_{\rm S}$
from variations in the PDF parameters.  The total uncertainty is the
sum in quadrature of the individual contributions.}
\begin{center}
\begin{tabular}{ccc} \hline\hline
Source & Positive variation & Negative variation \\ \hline
\multicolumn{1}{l}{Signal} \\
$\mes$     & $+0.11$ & $-0.62$ \\
$\de$      & $+0.39$ & $-0.37$ \\
${\cal F}$ & $+0.03$ & $-0.03$ \\
\multicolumn{1}{l}{Background} \\
$\mes$     & $+0.32$ & $-0.30$ \\
$\de$      & $+0.01$ & $-0.01$ \\
${\cal F}$ & $+0.86$ & $-0.85$ \\ \hline
Total      & $+1.00$  & $-1.16$ \\ \hline\hline
\end{tabular}
\end{center}
\label{tab:signalyield}
\end{table}

\begin{table}[!htb]
\caption{Summary of relative statistical and systematic uncertainties on the 
signal efficiency.  The total uncertainty is the sum in quadrature 
of the individual contributions.}
\begin{center}
\begin{tabular}{lc} \hline\hline
Source           & Uncertainty ($\%$) \\ \hline
Statistical      & $7.7$ \\
Tracking         & $1.6$ \\
Vertex Quality   & $2.5$ \\
Proton Selection & $3.0$ \\
DIRC Acceptance  & $1.0$ \\
$\cos{\theta_S}$ & $5.0$ \\ \hline
Total            & $10.2$ \\ \hline\hline
\end{tabular}
\end{center}
\label{tab:efficiency}
\end{table}

We calculate the $90\%$ C.L. upper limit on the branching fraction by increasing 
$N_{\rm S}^{\rm UL}$ by the total systematic error on the signal yield, and by 
decreasing the efficiency and number of $\BB$ events by their respective total 
uncertainties.  We find the flavor-averaged branching fraction 
$\BR(\Bz\to p\bar{p}) < 2.7\times 10^{-7}$ at the $90\%$ C.L.  This result 
improves the previous limit~\cite{belleppbar} by more than a factor of four.

In summary, we have performed a search for the decay $\Bz\to p\bar{p}$ in a 
sample of $88$ million $\BB$ events.  We find no evidence for a signal and set 
an upper limit on the branching fraction at $2.7\times 10^{-7}$.  This result 
rules out the calculation in~\cite{ref:QCDsumrule} based on QCD sum rules, 
while it is consistent with a recent calculation using the 
pole model~\cite{ref:PoleModel}, and with simple scaling of the measured
branching fraction for the decay $\Bz\to\Lambda_c^- p$.  

\label{sec:Acknowledgments}

We are grateful for the 
extraordinary contributions of our \pep2\ colleagues in
achieving the excellent luminosity and machine conditions
that have made this work possible.
The success of this project also relies critically on the 
expertise and dedication of the computing organizations that 
support \babar.
The collaborating institutions wish to thank 
SLAC for its support and the kind hospitality extended to them. 
This work is supported by the
US Department of Energy
and National Science Foundation, the
Natural Sciences and Engineering Research Council (Canada),
Institute of High Energy Physics (China), the
Commissariat \`a l'Energie Atomique and
Institut National de Physique Nucl\'eaire et de Physique des Particules
(France), the
Bundesministerium f\"ur Bildung und Forschung and
Deutsche Forschungsgemeinschaft
(Germany), the
Istituto Nazionale di Fisica Nucleare (Italy),
the Foundation for Fundamental Research on Matter (The Netherlands),
the Research Council of Norway, the
Ministry of Science and Technology of the Russian Federation, and the
Particle Physics and Astronomy Research Council (United Kingdom). 
Individuals have received support from 
the A. P. Sloan Foundation, 
the Research Corporation,
and the Alexander von Humboldt Foundation.


\begin{thebibliography}{99}
\bibitem{ref:cc}
Unless otherwise noted, the charge-conjugate mode is implied throughout
this paper, and the term ``proton'' refers to both protons and antiprotons.
\end{thebibliography}
\end{document}